\documentclass[useAMS,usenatbib]{mn2e}
\usepackage{graphicx} 

\title[2nd AGN in HE0450-2958]
{Near-IR observations of the HE0450-2958 system: discovery of a second AGN?
\thanks{Based on observations made with the ESO Very Large Telescope  at ESO Paranal Observatory, Chile, under program IDs 076.B-0693(C), 276.B-5011 and NASA/ESA HST program 10797 during cycle 15.}} 
\author[G. Letawe et al.]
{G. Letawe,$^1$\thanks{E-mail: gletawe@ulg.ac.be}
P. Magain,$^{1}$ V. Chantry,$^{1}$, Y. Letawe,$^{1}$\\
$^{1}$ Institut d'Astrophysique et  G\'eophysique, Universit\' e de
  Li\`ege, All\'ee du 6 Ao\^ut, 17, Sart Tilman (Bat. B5C), B-4000 Li\`ege,
Belgium}
\pagerange{\pageref{firstpage}--\pageref{lastpage}} \pubyear{2009}
\begin{document}
\maketitle
\label{firstpage}
\begin{abstract}
The QSO HE0450-2958 was brought to the front scene by the non-detection of its host galaxy and strong upper limits on the latter's luminosity. The QSO is also a powerful infrared emitter, in gravitational interaction with a strongly  distorted  UltraLuminous InfraRed companion galaxy.  We investigate the properties of the companion galaxy, through new near- and mid-infrared observations of the system obtained with NICMOS onboard HST, ISAAC and VISIR on the ESO VLT.  The companion galaxy is found to harbour a point source revealed only in the infrared, in what appears as a hole or dark patch in the optical images.   Various hypotheses on the nature of this point source are analyzed and it is found that the only plausible one is that it is a strongly reddened AGN hidden behind a thick dust cloud. The hypothesis that the QSO supermassive black hole might have been ejected from the companion galaxy in the course of a galactic collision involving 3-body black holes interaction is also reviewed, on the basis of this new insight on a definitely complex system. 
\end{abstract}
\section{Introduction}
In a previous study of the host galaxies of bright low-redshift QSOs \citep{letawe07} combining VLT spectra and HST/ACS images, we discovered a particularly interesting case, namely HE0450-2958,  which no host galaxy could be detected for \citep{mag05}.  From these data, we could set an upper limit on its host brightness, implying that for a QSO of that magnitude it was underluminous by at least a factor of six. This upper limit was confirmed by \citet{kim07}.

\begin{figure}
\center
\includegraphics[width=8.5cm]{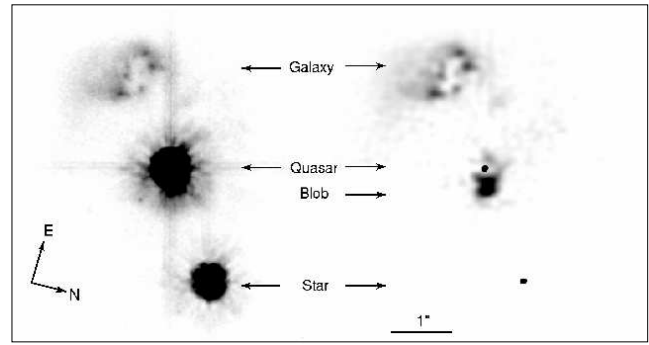}
\caption{Optical HST/ACS image of the quasar and surrounding objects, prior and after deconvolution.}
\label{acs}
\end{figure}
The QSO itself lies in a rather complex environment (Fig.~\ref{acs}), at 1.5arcsec (6.5kpc) NW from a strongly distorted companion galaxy. This system was also classified as Ultra Luminous Infrared Galaxy (ULIRG), from the IRAS survey \citep{degrijp}. Just besides the QSO, in the direction opposite to the companion galaxy, we discovered a compact emission region whose spectrum shows highly ionized gas with no trace of continuum light.  This gas is most probably ionized by the QSO itself.  An unrelated foreground star is also found 1.7 arcsec NW of the QSO.

Since the publication of our results \citep{mag05}, several papers appeared, proposing different explanations to our discovery. Two studies, \citet{hoffman} and \citet{haenelt} suggested that this system could result from the ejection of a supermassive black hole (SMBH) from a galaxy merger (a phenomenon predicted but not observed yet,  except a probable hint in \cite{komossa}).  If the quasar was ejected by radiation recoil, we would expect the nucleus of the companion to be empty of BHs.  If it was ejected in a 3-body encounter, then there should be another SMBH in the distorted companion galaxy, while none is observed in the optical (the presence of an inactive SMBH in such a violently interacting system seems unlikely).
On the other hand, \citet{merrit}  note that HE0450-2958, while being a Type 1 QSO, has comparatively narrow emission lines. They suggest that it could be a high luminosity analog to the so-called Narrow-Line Seyfert 1 Galaxies (NLS1s, \citealt{deo}), which are characterized by rather low-mass supermassive black holes (SMBHs) with high accretion rates (possibly above the Eddington limit), as confirmed by X-ray analyses (Zhou et al. 2007).  In such a case, if the tight empirical relation between the mass of the SMBH and the mass of the host galaxy bulge holds (e.g. \citealt{marconi}), one would expect a lower luminosity host, possibly just below our detection limit. 

From radio observations, \citet{feain}  claimed to have found evidence of star formation close to the quasar position, first hint of a host galaxy. However, the same authors, in a recent  publication \citep{papad}, have drastically changed their conclusions. They find CO emission (and thus molecular gas and star formation) in the companion galaxy, but not around the quasar. Their estimates confirm the companion galaxy as a ULIRG, and state that if any host galaxy exists, it has to be dominated by an old stellar population or it would have been detected.

A deeper spectroscopic  investigation of the whole system and its surrounding in the optical range (VIMOS/IFU and FORS/MXU from VLT) was also reported in a  previous paper \citep{let08}. It principally showed that: (1) the system is embedded in gas ionized by AGN and shocks, weakening the possibilty of a globally dust enshrouded host; (2) the dynamics of gas and stars are not related, witnessing perturbations by galaxy interaction; and (3) still no host galaxy could be detected.

The simplest explanation for the non-detection of the host is that it is devoid of a young stellar population component.  It might be just too faint in the HST/ACS F606W and the diverse optical VLT bandpasses we used and thus have escaped detection.  Infrared (IR) observations are thus the key to detect older stellar content and to investigate the presence of dust in this system.

Several additional  near-IR and mid-IR observations of the system have thus been achieved and are presented here. While the limits of detection on the host galaxy in all available wavebands and their consequences are presented in \cite{knud09},  discussing the tentative detection of a non-concentric part of the host and favouring the hypothesis of a NLS1-equivalent host just below the previous detection limits, we focus here on the companion galaxy. 

\section{Observations}
 HST/NICMOS and VLT/VISIR observations presented below are described in detail in \cite{knud09}, we just present here a summary and refer to this paper for a detailed description of reduction and analysis.
\subsection{HST/NICMOS}
We observed HE0450-2958 with the camera 2 of HST/NICMOS (Near Infrared Camera and Multi-Object Spectrometer), through the wide filter F160W, corresponding to the H band, on July 2006 and July 2007 (program 10797, PI:K. Jahnke). Data transmission problems and failure of guide star acquisition have poored the resulting observations, moreover splitting the run in two sets of different orientations. The PSF star planned to be observed also partially suffered from these bad transmission problems, making it difficult to use correctly. The images have a spatial scale of 0.0756\arcsec/pixel covering a field of view (FOV)  of 19.2$\times$19.2\arcsec, in which the HE0450-2958 is the only system detected. Total integration time on target was $2\times1280$s, and the reduction was performed with Pyraf tasks.

\subsection{VLT/ISAAC}
\begin{figure}
\includegraphics[width=8cm]{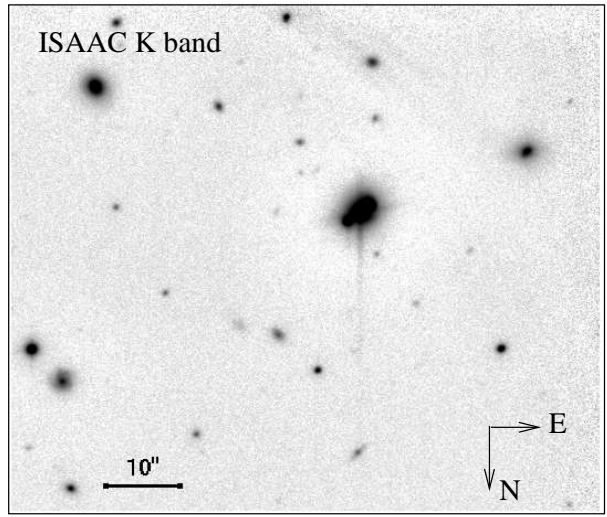}
\caption{Part of the neighbourhood of the system from the ISAAC K band image, after combination of all exposures (logarithmic intensity scale).}
\label{is_field}
\end{figure}

We used VLT/ISAAC to obtain deep observations in the K band (ESO program 076.B-0693(C)), in April 2006. The spatial resolution of 0.148\arcsec/pixel covers a field of 2.7$\times$2.7\arcmin, including the direct neighbourhood of the system (Fig.~\ref{is_field}). Total integration time is 2.2 hours, but saturation was reached for the quasar in individual subexposures for 2/3 of the observations. The reduction was performed with Pyraf tasks.
 
\subsection{VLT/VISIR}
Observations in the mid-IR were obtained with VLT/VISIR, tracking for the presence of dust, through the  PAH2 filter centered at 11.3 $\mu$m (ESO program 276.B-5011(A)). Total exposure time was 1623s, with a pixel scale of 0.075\arcsec, for a total FOV of 19.2$\times$19.2\arcsec. There is only one source detected in the field. This source appears point-like with a FWHM=0.35\arcsec, which is exactly what is expected for a diffraction limited observation of a point  source with VISIR. The observed flux is 62.5mJy, and we can reject the presence of a second object in the field down to a  least 3 mJy at the 5-sigma level. The position of this point source,  after correction of a systematic offset detected in the VISIR pointing (\cite{knud09}),  makes it compatible with the quasar position, which is thus probably the mid-IR emitter.

\section{Image analysis}
\label{anal}

We treated both NICMOS and ISAAC observations with the MCS deconvolution routines \citep{mcs}, which allows to improve the resolution of the observation while conforming to the sampling theorem and separating point sources from diffuse ones. Its power relies on the ability to construct reliable PSFs with sufficient accuracy. The HE0450-2958 system is a peculiar case, in the sense that three sources are close to each other and partially blended (a foreground star, the quasar and the companion galaxy, see Figs.~\ref{nicdec} and \ref{isdec}). In both NICMOS and ISAAC cases, we observed an  isolated star in order to constrain the PSF, but their usefulness was hampered by too much variation in the field of ISAAC and spectral energy distribution mismatch in the NICMOS filter. We thus adapted our procedures to extract a maximum information on the PSF from both the QSO and the  star. In both cases we based the deconvolution on the iterative procedure described in \cite{chantry}.

\subsection{NICMOS F160W images}
\begin{figure*}
\includegraphics[width=16cm]{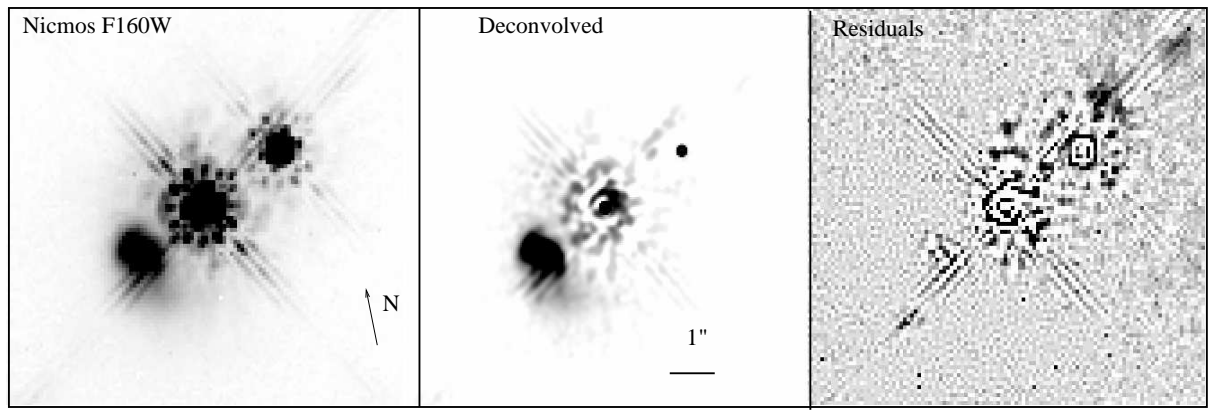}
\caption{NICMOS observation of the system (left) and an example of  deconvolution with the MCS routines (middle), where point sources are represented by the black dots of FWHM=2 pix. Associated residuals (model - data) are shown in the right panel.}
\label{nicdec}
\end{figure*}

 In addition to the treatement presented in \cite{knud09}, we made several tests to improve the PSF, which is mandatory for appropriate point source removal. An illustration is given in   Fig.~ \ref{nicdec}.
 We first built a PSF based on stars only (the separately observed PSF star and/or the foreground star). The deconvolution with these PSFs is good for the neighbouring star, but  not for the QSO. The residuals (difference between model and observation) show strong ringlike structures, revealing that the width of the PSF is not adapted. This indicates that the QSO spectral distribution is too different from the stellar ones, even if the separate PSF star was chosen for having a SED comparable with typical QSO ones.

To correct the PSF for the difference in spectral distribution, we use the Tinytim software, developed  to model HST PSFs as a function of wavelength.
 We build a tinytim PSF with a blackbody distribution at 5700K
accounting for the star and another one adapted to the quasar, the latter being a
combination of a power law (contribution of the accretion disk) and cooler black body (dust torus). We compute the difference of these Tinytim PSFs and add it to the star-based PSFs. The resulting modified PSF is used for deconvolution of the whole system.
The best results are obtained with a rather hard power law ($f_{\nu}\sim\nu^1$) for 70\% of the flux and a black body at 1000K for the remaining 30\%, making the internal ring structure in the residuals of the deconvolution almost disappear. These tests show that it is mandatory to include a red component in the spectral distribution of the quasar. Similar results are obtained when adding a compact gaussian contribution to the point source, broadening it as a redder SED would.

But even with such a modified PSF, the residuals in the quasar region are not yet acceptable (see Fig.~\ref{nicdec}, right). Moreover, the structure in the inner part of the background light under the quasar point source found by deconvolution  is different in the 2006 observations and 2007 reobservations, reinforcing our conclusion that what is observed is not real but just artefacts caused by the use of an improper PSF. The orientation of these observations are slightly different, so no simple simultaneous deconvolution is possible. The only similarities in both background images are found at larger scale ($\sim$ 1\arcsec around the quasar), and have exactly the same orientation as the typical NICMOS  square-like PSF structure, being thus associated to an unadapted PSF rather than to a putative host (see Fig.~\ref{nicdec}, middle).

 With the poor matching of the PSFs, the residuals close to the QSO position are compatible with deconvolution artefacts.  In summary, the main  conclusions of these tests are that (1) the QSO SED is significantly redder than typical (2) no co-centered host can be detected. More quantitative estimates on the host galaxy upper limits are presented in \cite{knud09}.

\subsection{Isaac K images}
\begin{figure*}
\includegraphics[width=16cm]{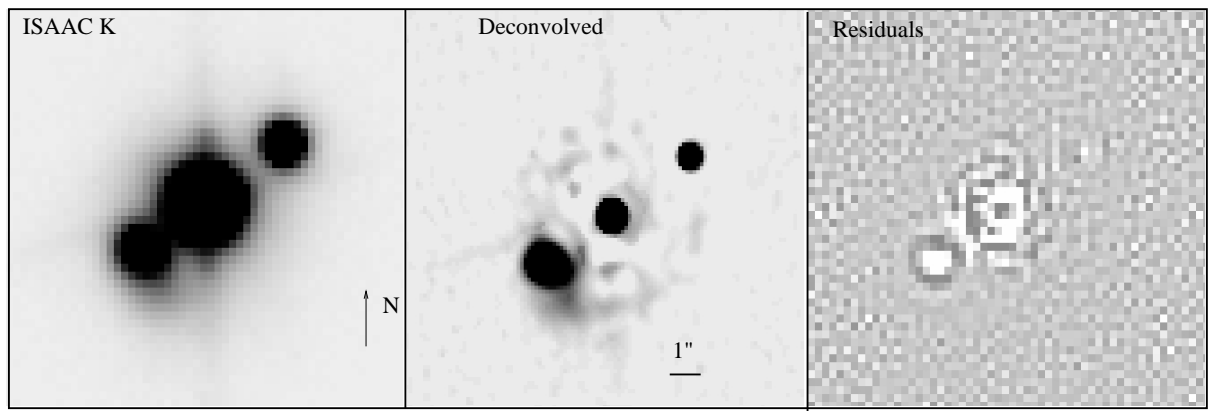}
\caption{HE0450-2958 system through the K band with VLT/ISAAC prior (left) and after (middle)  deconvolution, and associated residuals (right).}
\label{isdec}
\end{figure*}

The MCS deconvolution routines were applied only to the unsaturated exposures (i.e. a total of 43min  exposure time), allowing tentative removal of the point sources. 
No star bright enough for PSF construction is found in the vicinity of the QSO, except the very close one. Brighter stars are found further away, but it appears that the variations of the PSF through the ISAAC field are too important to allow using them to construct appropriate PSFs. We thus used the QSO itself and the companion star to constrain the PSF. As for the NICMOS observations, no appropriate PSF could be used to remove properly the quasar contribution, the SED mismatch between the quasar and the star adding to the loss of linearity of the detector almost reached in these observations. Results are presented in Fig.~\ref{isdec}, where we show the bad residuals on the right panel. 

But whereas no useful constraint can be set on the host galaxy from these observations, the companion galaxy analysis can shed light on our understanding of the system.

\subsection{The companion galaxy}
\begin{figure}
\center
\includegraphics[width=8cm]{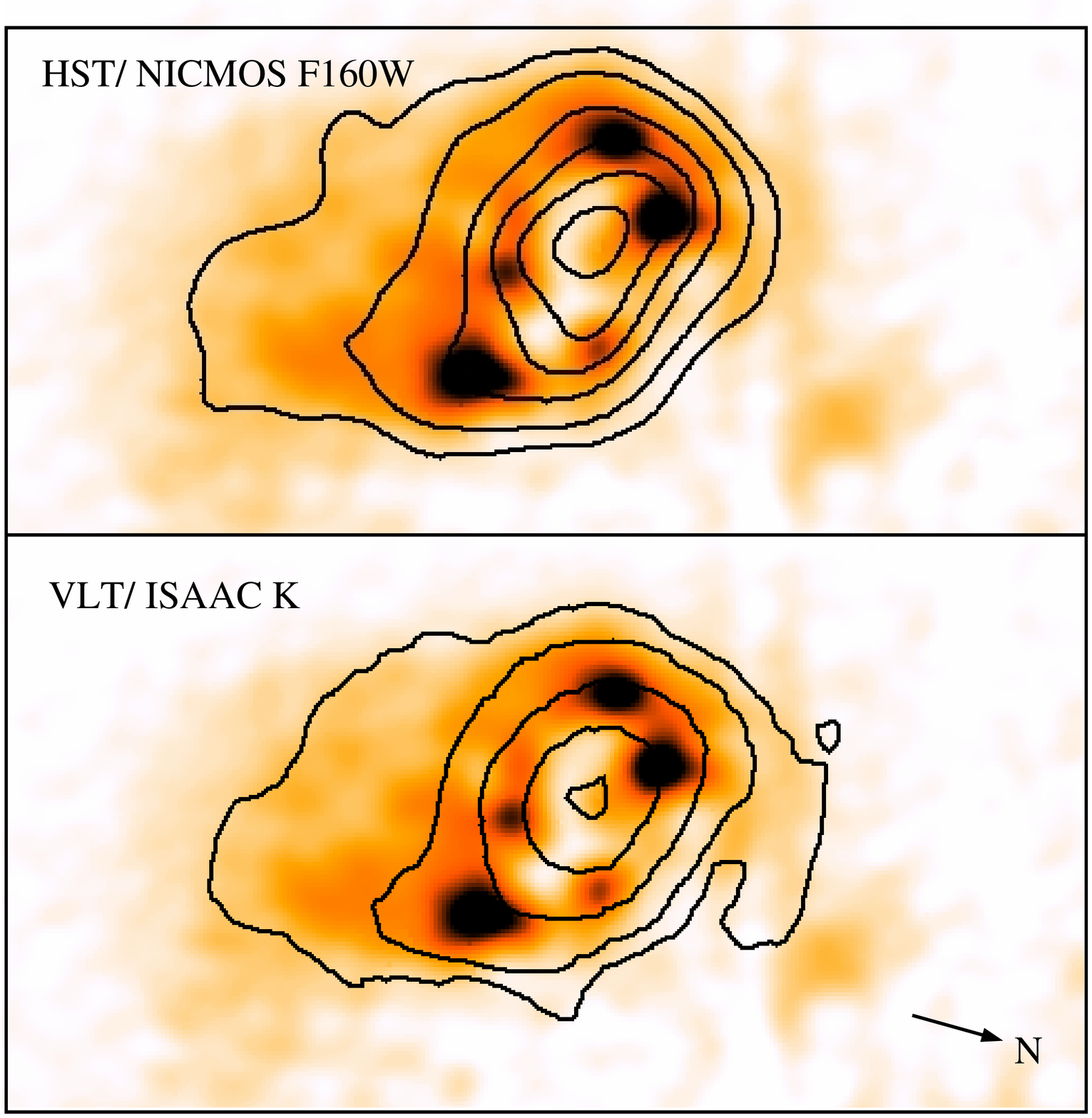}
\caption{The companion galaxy from HST/ACS, on which we overplot the contours of near-IR data (NICMOS F160W in the top panel and ISAAC K at the bottom), after the main QSO point source removal with the MCS algorithm.}
\label{irgal}
\end{figure}

\begin{figure}
\center
\includegraphics[width=8cm]{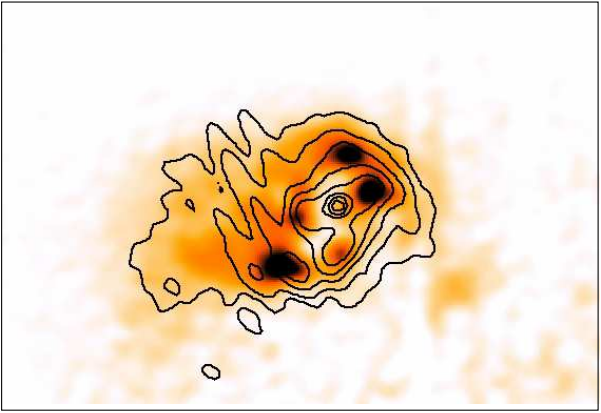}
\caption{The companion galaxy from HST/ACS, on which we overplot the contours of near-IR data Nicmos F160W after deconvolution, including a point source. The  wavy structure at the top left is caused by residual spikes from the luminous quasar that are not properly removed because of PSF mismatch.}
\label{ir_sp}
\end{figure}

\begin{figure}
\includegraphics[width=7.5cm]{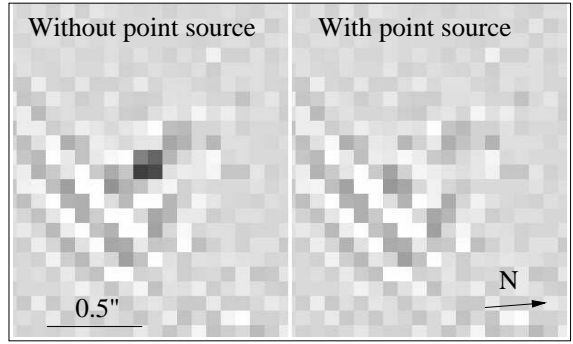}
\caption{Illustration of the improvement of the residuals by the addition of a point source  in the companion galaxy, in the NICMOS observations. Left is without, right is with a point source added to the deconvolution routines.  The nearly diagonal linear structures are due to an imperfect modeling of the spikes in the PSF of the bright QSO.}
\label{compresi}
\end{figure}

The near-IR HST image of the companion galaxy (Fig.~\ref{irgal}) shows that the most intense near-IR emission comes from the region which appears as a hole in the optical.  This means that there is indeed no hole, but rather a huge dust cloud (about 5 kpc in length) which strongly absorbs the optical light but is more transparent to IR radiation. Trying to understand the nature of this emission, we noted that the best deconvolution residuals are obtained when adding a point source at the location of this near-IR peak (Fig.~\ref{ir_sp}).  An illustration of the improvement of the residuals (observation minus model, divided by the standard deviation) when a point source is added, is given in Fig.~\ref{compresi}. More quantitatively, the reduced $\chi^2$ on the companion galaxy is reduced by a factor $\sim$ 5, from 5.7 to 1.1, when computed on a 10 $\times$ 10 pixels area around the putative point source. The deconvolution of this part of the system is quite insensitive to the PSF SED mismatch, as this point source is of a much lower intensity. This near-IR peak is spatially unresolved both with HST and the ESO VLT. These data are thus compatible with the presence of a point source lying in (or behind) the dust cloud, with a flux of 5.6$\pm$1.5 × 10$^{-6}$ Jy at 1.6 $\mu m$ and 34.8$\pm$10.8 × 10$^{-6}$ Jy at 2.2 $\mu m$, where the error bars denote the dispersion of the point source fluxes obtained from the different exposures and observing runs in each band.  The question which then arises is: what is the nature of this compact IR source?

\section{The nature of the unresolved IR source}

\subsection{Emitting dust cloud?}
A first assumption is that the unresolved  IR source could correspond to a compact cloud of hot dust, emitting in the near-IR.  In this case, using the Planck distribution for a blackbody, the 1.6 to 2.2 $\mu m$ flux ratio implies a temperature of 1000$\pm$100~K, and the non detection of this source at 11.3 $\mu m$ imposes it  to be above 730$\pm$25 K.  More precisely, the data imply that no significant part of the dust cloud would be found at temperatures below 700 K.  This appears unrealistically high, as dust heated, e.g. by the UV radiation of young stars, generally has temperatures below 100 K \citep{hunt}.  Very hot dust has only been found in the immediate vicinity of the centre of active galaxies (quasars or Seyferts).  Even in such cases, the highest temperatures measured do not exceed 600 K \citep{rouan}.   The explanation of the near-IR emission in terms of hot dust can thus be safely ruled out.

\subsection{Compact starburst region?}
Another hypothesis is that it would correspond to a bright source located inside or behind the dust cloud, and whose radiation would be dimmed and reddened by dust absorption.  Let us first assume that it corresponds to a compact starburst region. We also assume that the measured near-IR flux ratio (1.6 to 2.2 $\mu m$), in comparison with the expected value for a starburst, can reveal the amount of reddening.
The observed fluxes, including Vega zeropoints, give the colour
\begin{equation}
(F160W-K)_{obs}=2.59 \pm 0.27
\end{equation} 
Using a synthetic starburst (SB) model from \citet{dopita}, we extract the fluxes in the NICMOS F160W and ISAAC K filters, after having shifted them to the restframe of HE0450-2958 (z=0.258), giving the expected colour for a starburst region at the redshift of interest. 
\begin{equation}
(F160W-K)_{SB,z=0.285}=0.606
\end{equation} 
 Using an average dust reddening law (from Mathis, as found in \citealt{kox}), this can be translated into  a value for the dust absorption $A_V$ in the optical, assuming that the extinction in F160W is roughly equal to $A_H$:
\begin{eqnarray*}
A_{H}-A_K&= &\\(F160W&-&K)_{obs}- (F160W-K)_{SB,z=0.285}\\
&=&1.99 \\
\Rightarrow A_V&=& (A_{H} - A_K )/0.106=18.75\pm2.6\\
\end{eqnarray*} 

 The intrinsic (i.e. unreddened by dust) absolute visual magnitude $M_V$ is derived on the basis of the observed flux for the point source in ISAAC converted to an apparent magnitude $H=18.64$. $A_H= 0.176 implies A_V  =3.3$ with the adopted extinction law, and given the color index $V-H=-1.56$ as deduced from the synthetic spectra, we obtain V=16.9 and an absolute magnitude $M_V$ of $-24.4$ ($\pm$2.8) given the distances in our adopted cosmology ($h_0=0.65$, $\Omega_m=0.3$ and $\Omega_{\Lambda}=0.7$).

  Such extreme brightness is found only for the most massive galaxies in rich clusters or for quasars, but certainly not for starburst regions in galaxies such as the one considered here.  As an example, the well-known Antennae galaxies (NGC4038/4039), which experience strong collision-induced starburst, have a total absolute magnitude of $M_V \sim -18$, i.e. a factor 250 fainter than our aforementioned estimate.  The hypothesis of a compact starburst region strongly reddened by dust can thus also be ruled out.

\subsection{Hidden AGN?}
The remaining hypothesis is that the compact near-IR source is a strongly reddened active galactic nucleus (AGN).  In this case, we can derive an estimate of the dust absorption, using a similar method as for the starburst region, but with a spectrum typical of an AGN, i.e. a power law $F_\nu \sim \nu^\alpha$, with an exponent $\alpha$ ranging from $-1$ to $+0.5$, encompassing the range found in quasar spectra.  We obtain A$_V$ between 16 and 21 ($\pm$3) and an absolute magnitude  $M_V$ ranging from $-23$ to $-25.7$ ($\pm$ 2.8), depending  on the adopted power law exponent.  To fix a value, we choose $\alpha=-0.3$ as an average quasar power law, giving $A_V=18.64$ and $M_V=-24.22\pm2.8$. These absolute magnitudes correspond quite well to  bright AGN (the limit for QSO/Seyfert separation is usually set at $M_V = -23$).

\section{Interpretation}
\begin{figure}
\includegraphics[width=8cm]{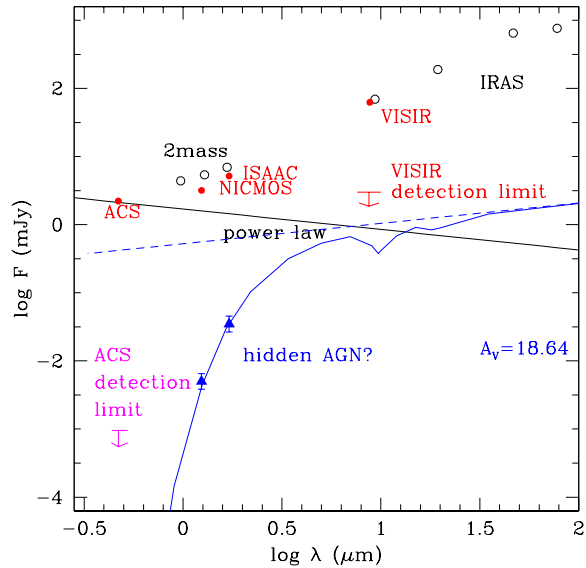}
\caption{Spectral energy distribution of the QSO. Open circles refer to previous observations with low spatial resolution, including all intervening objects (galaxy+star+quasar), from 2mass and IRAS surveys. Filled circles correspond to the QSO light only, measured on the present observations (HST/ACS, HST/NICMOS, VLT/ISAAC and VLT/VISIR). The main QSO power law (straight solid black line) was evaluated from VLT/FORS optical spectra, and corresponds to the expected contribution from the accretion disk. Triangles refer to the near-IR point source emission in the companion galaxy. The dashed line corresponds to an average QSO power law, and the solid curve to the same power law, after reddening by dust ($A_V=18.6$) has been applied to fit the near-IR data. The arrows indicate the upper limits at the position of the near-IR point source in the companion galaxy, as derived from the HST/ACS and VLT/VISIR observations.}
\label{SED}
\end{figure}

As shown in the SED presented in Fig.~\ref{SED}, the hypothesis of an AGN strongly obscured by dust is compatible with the non detection of the point source both in the HST/ACS optical images and in the VLT/VISIR mid-IR observations.

 The ACS detection limit was set by adding an artificial point source to the original image and dimming it until it could not be detected anymore, and corresponds to m$_{F606W}=24.2$. In fact, A$_V=3$ would be sufficient to hide this point source in the optical.   

The upper limit derived from VISIR data is based on the non-detection of any other source in the field down to a 3$\sigma$ detection level. 
As proposed in \citet{glikman}, a typical quasar SED in the near-infrared can be reproduced by a power law plus a black body of T$\sim$1250K accounting for the hot dust. On Fig.~\ref{SED} only the power law component is presented. 
 
The VISIR detection limit  is a factor 3 above the predicted flux of an average AGN power law giving the observed values in the NIR after reddening.  This leaves only rather little room for additional contributions from dust heated by the AGN itself. However, direct comparison with emission from hot dust around the main QSO is avoided, as this one is an extreme IR emittor. 

To quantify the contribution of dust  in the VISIR band, we refer to \citet{haas} and his mid-IR observation of a 17 PG quasars sample. He finds an average difference in log F$_\nu$ between 1$\mu m$ and 10$\mu m$  of $\Delta log F_\nu= 1.1$. Adding this flux to the unreddened power law at 10 microns leads to an expected flux  (power law + dust) falling exactly on the detection limit. An AGN harbouring hot dust just below the average would thus not be detected on our VISIR observations.

  We thus find that the assumption of an AGN embedded in (or hidden behind) a thick dust cloud is compatible with all the available data, assuming its hot dust component is not too strong.  Converting the unreddened luminosity of this AGN to a black hole mass, according to \cite{floyd04}, leads to $M_{\rm{BH}}\sim 2\cdot 10^8 M_{\odot} $(with an AGN accreting at 50\% of the Eddigton limit).
\\
 The presence of an AGN in the ULIRG is not surprizing. Following \cite{sandersmirab}, who studied a large sample of ULIRGs, about one third of them are harbouring an active SMBH, while 95\% are found in mergers, which is also the case in HE0450-2958 at first sight. Only the projected separation between the two components of the HE0450-2958 merger ($\sim$ 7 kpc) is found larger than the mean separation in their sample (1.2 kpc).  But there is little doubt that HE0450-2958 is rather far from being an `average' object of any kind.

 While active nuclei in ULIRG are expected, the observation of merging galaxies both containing AGN is scarce. In the literature one finds evidence for two ULIRG systems where at least one of the AGN is obscured and only revealed by X-ray observations: Arp 229 in \cite{ballo} and NGC 6240 in \cite{komossa03}. One more unobscured system, LBQS 0103-2753, not ULIRG, is observed with 3.5kpc separation between the QSOs by \cite{junk}. All three might be illustrative of the merging scenario for the formation and evolution of AGN and galaxies, as the final stage of the presented systems is supposed to be the merging of both SMBH and galaxies.

The presence of this second SMBH  reintroduces the 3-body BH ejection hypothesis as a possible explanation of this exotic system. As presented in \citet{haenelt} and \citet{hoffman}, a merger of two galaxies, with one harbouring a binary black hole  resulting from a previous merger, and the second a third SMBH, could eject this third BH at high speed while the binary would suffer a slight recoil.  The hypothesis of gravitational radiation recoil, where the resulting SMBH binary formed in the merging of two  galaxies is ejected, does not appear compatible with our present discovery, as no SMBH would be left in the galaxy.  
The main quasar would correspond to the lighter black hole  (M$_{\rm{BH}}\sim4\cdot 10^7M_{\odot}$), ejected with its accretion disk and dust torus (explaining the powerful optical spectra and intense near- and mid-IR emissions as well as redder-than-expected NICMOS F160W quasar SED), while the binary (M$_{\rm{BH}}\sim2\cdot 10^8M_{\odot}$) would still be in the galaxy, with a normal dust torus, but placed behind a thick dust cloud that would redden it and hide it in the optical.

 Several observational arguments are compatible with this scenario, while some points still remain to be elucidated. We start by listing the data compatible with the  3-body BH ejection scenario:
\begin{itemize}
 \item The mass ratio between the putative binary and the isolated quasar is estimated at 5:1. This ratio is compatible with ejection in a 3-body interaction, according to simulations made by \cite{hoffman07}.  Moreover such a binary mass implies a magnitude similar to the magnitude of the whole ULIRG itself, and thus would have been detected without the presence of large dust cloud.
\item The analysis of the H$\alpha$ line from the main QSO reflected on moving material between the QSO and the ULIRG made by \cite{yan}, points to a reflecting material moving towards the ULIRG and away from us, at about 45$^o$. The isolated (maybe ejected) QSO would then be in the foreground, and the heaviest AGN in the background.  Some recoil would be compatible with its location behind most of the dust found in the merging galaxies.
\item The distance between the quasar and the galaxy is quite large (at least 7 kpc), but compatible with the simulations presented in \citet{hoffman}, if the quasar is close to its largest distance, before turning back to the ULIRG. This is also compatible with the small radial velocity difference measured between the two components.
\item In a recent publication, \citet{komossa} present what might be the first observation of a recoiling black hole, SDSSJ0927+2943, whose expected recoiling signature is the large shift ($\sim$ 2650 km$  s^{-1}$) between broad and narrow nuclear emission lines, as the SMBH takes its accretion disk away with him and leaves the narrow line emitting gas behind. HE0450-2958 could be another case, still more exotic. The narrow lines seen in the quasar might just arise from the blob, a huge gas cloud illuminated  in situ by the ejected quasar. The presence of such an isolated cloud (outside a galaxy) is not surprizing in a merger of galaxies. Others have indeed been detected by \cite{let08}. Moreover, as noticed by \cite{merrit}, the narrow line region (NLR) of this AGN cannot have been ejected with it because of its large spatial extension. The scrutiny of the [OIII] emission lines from optical spectra and HST/ACS imaging shows that the nuclear narrow emission lines are non negligible, but can be explained if the cloud of ionized gas overlaps the line of sight to the QSO, or if this blob becomes the NLR as the QSO passes through it. 

\end{itemize}
Some observational data still need to be explained:
\begin{itemize}

\item The mass reservoir provided by the blob, estimated by $M_{\rm{NLR}}=7\cdot 10^5\frac{L(H\beta)\cdot 10^{41}{\rm erg.s}^{-1}}{{\rm n}_{e^-} 10^3{\rm cm^3}}M_{\odot}=3.4\cdot 10^6 M_{\odot}$ (\cite{peterson}), is not sufficient for AGN fueling with the observed accretion rate of 1.4M$_{\odot}$/yr as estimated in \cite{knud09} during its journey of 2$\cdot 10^7$yr since ejection (\cite{hoffman}). The quasar thus needs to have taken its feeding material with it during ejection. With an assumed constant accretion rate, these values give an amount of $2.8\cdot 10^7$M$_{\odot}$ accreted along its path, that would have doubled the black hole mass since ejection. Further 3-body interaction simulations are required to state if these mass estimates are compatible with the ejection scenario.  A way to alleviate this problem would be to assume that the activity started more recently, when the ejected BH reached the gas cloud.
 \item The QSO is a strong IR emitter (to ULIRG level, \cite{knud09}). It thus has to be ejected not only with sufficient matter to fuel its activity during its journey, but also with its dust torus to explain its IR emission. Once again numerical 3-body dynamical simulations with our new mass estimates could indicate if some appropriate input parameters can reproduce the observed configuration.
 \item While the narrow [OIII] emission lines could be originating from the blob only, we cannot exclude a nuclear contribution, that would be more difficult to justify in the framework of an ejection scenario.
\end{itemize}

\section{Conclusion}
Several attempts to detect the host galaxy of the QSO HE0450-2958 have given similar upper limits, without enabling to firmly exclude or confirm its existence so far. Even if a galaxy was finally detected around this AGN,  as partially suggested in \cite{knud09}, albeit on the basis of rather weak observational evidence, this system would remain a peculiar one in several features:  (i) an ULIRG (the companion galaxy)  associated to a QSO that does not belong to it, (ii) a particulary red QSO, while not obscured by dust, (iii) an important cloud of ionized gas being the  main emitting region close to the QSO.

Moreover, the presence of an AGN hidden behind a thick dust cloud in the companion galaxy  appears as the best interpretation of the detected near-IR compact source.  The presence of this AGN is a necessary (but not sufficient) condition for the SMBH ejection hypothesis to be viable, a scenario that does not require the presence of a host galaxy around HE0450-2958.  However, additional dynamical simulations appear necessary to further test this hypothesis on the light of the new data. In any case, deeper near-IR observations, with careful removal of appropriate PSFs, may allow to give a definite explanation to this puzzling system, if a host galaxy is detected.

\section*{Acknowledgements}
G.L. is a teaching assistant  supported by the University of Li\`ege
(Belgium). V.C. is a Research Fellow from the Belgian National Fund for Scientific Research (FNRS).  This work has been supported by ESA and the PPS Science Policy, Belgium, through PRODEX PEA 90312.

\label{lastpage}
\end{document}